\documentclass[prb,twocolumn,showpacs,amsmath,amssymb]{revtex4}
\usepackage{graphicx}
\usepackage{dcolumn}
\usepackage{bm}
\usepackage{subfigure}
\usepackage{epsfig}
\begin{document}


\title{Magnetic order in the $S=1/2$
two-dimensional molecular antiferromagnet, copper pyrazine
perchlorate Cu(Pz)$_{2}$(ClO$_{4}$)$_{2}$}

\author{T. Lancaster}
\email{t.lancaster1@physics.ox.ac.uk}
\author{S. J. Blundell}
\author{M. L. Brooks}
\author{P. J. Baker}
\affiliation{
Clarendon Laboratory, Oxford University Department of Physics, Parks
Road, Oxford, OX1 3PU, UK
}
\author{F. L. Pratt}
\affiliation{
ISIS Facility, Rutherford Appleton Laboratory, Chilton, 
Oxfordshire OX11 0QX, UK}
\author{J. L. Manson}
\author{M. M. Conner}
\affiliation{Department of Chemistry and Biochemistry,
Eastern Washington University,
Cheney, WA 99004, USA}
\author{F. Xiao}
\author{C. P. Landee}
\affiliation{Department of Physics, Clark
University, Worcester, Massachusetts 01610, USA}
\author{F. A. Chaves}
\author{S. Soriano} 
\altaffiliation{Present address: DIRO-UFF, Rio das Ostras, Rio de
  Janeiro, Brazil}
\author{M. A. Novak}
\affiliation{Instituto de F'sica, UFRJ, Rio de Janeiro 21945-970, Brazil}
\author{T. Papageorgiou}
\author{A. Bianchi} 
\altaffiliation{Present address: Department of Physics and Astronomy,
University of California, Irvine, CA 92697, USA}
\author{T. Herrmannsd\"{o}rfer}
\author{J. Wosnitza}
\affiliation{
Hochfeld-Magnetlabor Dresden (HLD), Forschungszentrum Dresden - Rossendorf,
D-01314 Dresden, Germany
}
\author{J. A. Schlueter}
\affiliation{Material Science Division, Argonne National Laboratory, 
Argonne, IL 60439, USA}

\date{\today}

\begin{abstract}
  We present an investigation of magnetic ordering in the two-dimensional
  $S=1/2$ quantum magnet Cu(Pz)$_{2}$(ClO$_{4}$)$_{2}$ using specific heat
  and zero field muon-spin relaxation ($\mu^+$SR). The magnetic
  contribution to the specific heat is consistent with an exchange
  strength of 17.7(3)~K. We find unambiguous evidence for a transition to
  a state of three-dimensional long range order below a critical
  temperature $T_{\mathrm{N}}=4.21(1)$~K using $\mu^+$SR even though there
  is no feature in the specific heat at that temperature.  The absence of
  a specific heat anomaly at $T_{\mathrm{N}}$ is consistent with recent
  theoretical predictions.  The ratio of $T_{\mathrm{N}}/J=0.24$
  corresponds to a ratio of intralayer to interlayer exchange constants of
  $|J'/J|=6.8 \times 10^{-4}$, indicative of excellent two-dimensional
  isolation.  The scaled magnetic specific heat of
  [Cu(Pz)$_2$(HF$_2$)]BF$_4$, a compound with an analogous structure, is
  essentially identical to that of Cu(Pz)$_2$(ClO$_4$)$_2$ although both
  differ slightly from the predicted value for an ideal 2D $S$=1/2
  Heisenberg antiferromagnet.
\end{abstract}

\pacs{75.50.Xx, 75.50.Ee, 76.75.+i}
\maketitle

\section{Introduction}

The properties of systems described by
the $S=1/2$ two-dimensional square lattice quantum Heisenberg
antiferromagnet (2DSLQHA) model\cite{manousakis}
remain
some of the most pressing problems in condensed matter physics.
The 2DSLQHA model is described by the Hamiltonian
\begin{equation}
H = \sum_{\langle ij \rangle} J \mathbf{S}_{i} \cdot \mathbf{S}_{j},
\end{equation}
where $J$ is the nearest neighbour in-plane superexchange
interaction. Although long range magnetic order (LRO) does not occur
within the model above zero temperature\cite{mermin}, layered
materials that are well described by this model inevitably possess
some degree of interlayer coupling (quantified by a coupling
constant $J'$) that leads to a crossover to a regime of
three-dimensional (3D) LRO at a non-zero N\'{e}el temperature
$T_{\mathrm{N}}$. The identification of $T_{\mathrm{N}}$ and hence
the ratio $k_{\mathrm{B}}T_{\mathrm{N}}/J$ allows the evaluation of
the extent to which a material may be described by the 2DSLQHA
model.

The measurement of $T_{\mathrm{N}}$ is often problematical in
anisotropic spin systems due to the reduced ordered moment that
typify these materials. In addition, the short-range order that
exists in the two-dimensional (2D) planes of a layered material
above a 3D transition reduces the effective number of degrees of
freedom involved in the transition, diminishing the expected anomaly
in specific heat\cite{sengupta}. Our recent study\cite{tom1} of the
quasi one-dimensional (1D) $S=1/2$ chain compound
CuPz(NO$_{3}$)$_{2}$ (where Pz is pyrazine (C$_{4}$H$_{4}$N$_{2}$))
has shown that implanted muons are uniquely sensitive to the
presence of magnetic order in quasi-1D materials. In this paper we
present an investigation of the magnetic properties of the
two-dimensional analogue of CuPz(NO$_{3}$)$_{2}$, namely copper
pyrazine perchlorate (Cu(Pz)$_{2}$(ClO$_{4}$)$_{2}$). This material
has been the subject of several studies aimed at elucidating its
properties \cite{haddad,darriet,turnbull,albrecht,choi,woodward},
although hitherto, no evidence of a magnetic transition has been
found. We have carried out detailed specific heat and muon-spin
relaxation measurements. The latter provide unambiguous evidence for
a transition to a state of 3D LRO. As expected from recent
theoretical studies of highly anisotropic spin systems, our specific
heat measurements show no anomaly at the critical temperature and
are shown to be in very good agreement with the recent predictions
of Monte Carlo simulations of the 2DSLQHA \cite{sengupta}.

Copper pyrazine perchlorate (Cu(Pz)$_{2}$(ClO$_{4}$)$_{2}$) has long
been held to be a good example of a 2DSLQHA\cite{darriet}. The
material is formed from layers consisting of rectangular arrays of
Cu$^{2+}$ ions bridged by pyrazine ligands, which provide the
intralayer superexchange\cite{woodward}. The 300~K structure of
Cu(Pz)$_{2}$(ClO$_{4}$)$_{2}$ has the Cu$^{2+}$ ions semicoordinated
with two disordered, crystallographically inequivalent,
 ClO$_{4}^{-}$ tetrahedra (one above and
one below the layer) with $C2/m$ symmetry\cite{darriet}. The
ClO$_{4}^{-}$ ions order at low temperatures, where the crystal
structure is reduced to $C2/c$, with two the sets of identical
pyrazine pairs tilted by 62.8$^{\circ}$ and 69.1$^{\circ}$ with
respect to the Cu-N coordination plane\cite{woodward}. This results
in a herring-bone configuration, rather than a square lattice.
Despite this distortion, magnetic susceptibility measurements fit
very well to the 2DSLQHA model\cite{albrecht,turnbull,woodward} with
$J=17.5$~K and $g=2.11$.

A second copper pyrazine compound, [Cu(Pz)$_2$(HF$_2$)]BF$_4$, has
recently been reported\cite{manson2006}. X-ray
measurements\cite{manson2006} show it to have a similar layered structure
to Cu(Pz)$_2$(ClO$_4$)$_2$, but with adjacent layers bridged by HF$_2^-$
ions. The intralayer exchange strength and 3D ordering temperature were
found to be 5.7~K and 1.54~K, respectively.

\section{Experimental details}
\subsection{Specific heat}
The heat capacity measurements on Cu(Pz)$_2$(ClO$_4$)$_2$ have been
carried out on four small crystals with total mass of 75.7 mg fixed on a
Cu sample holder containing a Constantan heater and a calibrated Cernox
thermometer with Apiezon-N grease. We used a home-made calorimeter
attached to a Janis He$^3$ refrigerator, with an automated system based on
the semi-adiabatic method. The contribution of the sample holder as well
as of the grease to the total heat capacity had been subtracted.  Specific
heat measurements for [Cu(Pz)$_2$(HF$_2$)]BF$_4$ have been reported
previously\cite{manson2006}.

\subsection{Muon-spin relaxation}
Zero field muon-spin relaxation (ZF $\mu^{+}$SR) measurements have
been made on a powder sample of Cu(Pz)$_{2}$(ClO$_{4}$)$_{2}$ using
the MuSR instrument at the ISIS facility, Rutherford Appleton
Laboratory, UK.
In a $\mu^{+}$SR experiment \cite{steve} spin-polarized
positive muons are stopped in a target sample, where the muon usually
occupies an interstitial position in the crystal.
The observed property in the experiment is the time evolution of the
muon spin polarization, the behaviour of which depends on the
local magnetic field at
the muon site.
Each muon decays, with an average
lifetime of 2.2~$\mu$s, into two neutrinos and a
positron, the
latter particle being
emitted preferentially along
the instantaneous direction of the muon spin.
Recording the time dependence of the positron emission directions
therefore allows the determination of
the spin-polarization of the ensemble of muons.
In our experiments positrons are
detected by detectors placed forward (F) and
backward (B) of the initial muon polarization direction.
Histograms $N_{\mathrm{F}}(t)$ and $N_{\mathrm{B}}(t)$ record the number of
positrons detected in the two detectors as a
function of time following
the muon implantation. The quantity of interest is
the decay
positron asymmetry function, defined as
\begin{equation}
A(t)=\frac{N_{\mathrm{F}}(t)-\alpha_{\mathrm{exp}} N_{\mathrm{B}}(t)}
{N_{\mathrm{F}}(t)+\alpha_{\mathrm{exp}} N_{\mathrm{B}}(t)} \, ,
\end{equation}
where $\alpha_{\mathrm{exp}}$ is an
experimental calibration constant. $A(t)$ is proportional to the
spin polarization of the muon ensemble.

For these measurements, an Oxford Instruments Variox cryostat and a
sorption pump cryostat were used
\cite{isismusr}, with a powder sample packed in a
silver foil packet (Ag thickness 25~$\mu$m)
and mounted on a silver backing plate.
Silver is used since it possesses only
a small
nuclear moment
and so minimizes any background depolarizing signal.

\section{Results}

\subsection{Specific heat}
The magnetic specific heat of the 2DSLQHA has been calculated using a
variety of techniques. Early efforts concentrated on high temperature
series expansions \cite{navarro} while more recent work has involved
quantum Monte Carlo simulations \cite{sengupta, makivic91}. These studies
reveal that the specific heat rises rapidly from zero at low temperature
to a rounded maximum (of magnitude 3.8 J/K$\cdot$mol) at a temperature
$T_{\mathrm{max}}$ of 0.65$J$ before decreasing at higher temperatures. A
rough approximation to the exchange strength is thus
$J=T_{\mathrm{max}}/0.65 =1.54~T_{\mathrm{max}}$. The presence of
interlayer interactions $J'$ induces a transition to long-range order and
causes the existence of a local maximum in the specific heat at $T_N$;
this maximum is undetectably small for ratios $|J'/J|<0.01$ but grows
rapidly as the ratio approaches unity \cite{sengupta}.

The magnetic contribution of the ideal 2DSLQHA ($J'=0$) can be
represented as a function of temperature as a ratio of polynomials
in powers of $T/J$ using the following equation
\begin{equation}
C_{\mathrm{mag}}=R\frac{\sum^{5}_{n=1}C_n(T/J)^n}{\sum^5_{n=1}D_n(T/J)^n}
\label{CmagEq}
\end{equation}
where $R$ is the gas constant and $C_n$, $D_n$ are coefficients obtained
by fitting data from Monte Carlo Simulation\cite{sengupta}. The
coefficients appear in Table~\ref{SHCOEF}. This form is useful for
comparing experimental data to the theoretical predictions and is valid
for $T/J>0.1$.

\begin{table}[h!]
   \begin{center}
     \begin{ruledtabular}
   \begin{tabular}{lll}  
$n$ &  $C_{n}$  & $D_{n}$ \\
\hline
1   &  $-$0.0036      &   1.86131  \\
2   &  0.26197     &    $-$11.51455  \\
3   &  $-$1.45326    &   28.45991   \\
4   &  3.06373    &  $-$32.81602     \\
5    &$-$0.38349    &18.6304     \\
   \end{tabular}
     \end{ruledtabular}
   \end{center}
   \caption{Fitting coefficients for pure 2DSLQHAF specific heat
     polynomials
     from Ref.~\onlinecite{sengupta}.\label{SHCOEF} }
\end{table}\

\begin{figure}
\epsfig{file=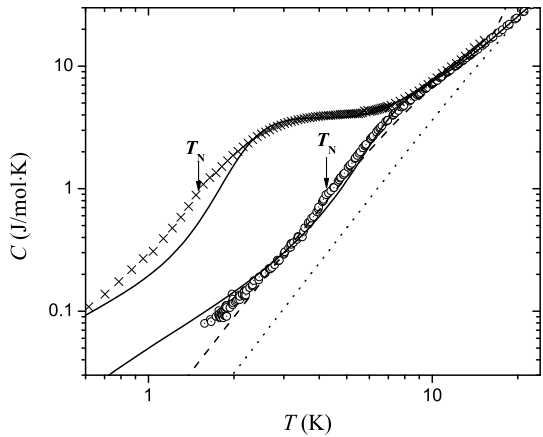,width=7.5cm} \caption{Specific heats of
Cu(Pz)$_2$(ClO$_4$)$_2$ ($\odot$) and [Cu(Pz)$_2$(HF$_2$)]BF$_4$
($\times$) as functions of temperature. The solid lines correspond
to the best-fit predictions for the specific heats based on the
parameters given in the text. The dashed lines represent the
estimated lattice specific heats for Cu(Pz)$_2$(ClO$_4$)$_2$ (short
dashes) and [Cu(Pz)$_2$(HF$_2$)]BF$_4$ (long dashes). The arrows
mark the temperatures of the magnetic ordering transitions for the
two compounds as determined by muon spin relaxation studies. Note
the absence of anomalies in the specific heats at those
temperatures.
  \label{totalSpecificheat}}
\end{figure}
The molar specific heat data for Cu(Pz)$_2$(ClO$_4$)$_2$,
represented as circles, are shown as a function of temperature in
Fig.~\ref{totalSpecificheat}. There is no low-temperature maximum in
the data but an inflection point occurs near 8~K, indicative of a
non-lattice contribution. In contrast, the specific heat of the
analogous 2D copper pyrazine compound, [Cu(Pz)$_2$(HF$_2$)]BF$_4$,
also shown in Fig.~\ref{totalSpecificheat}, shows a definite
low-temperature anomaly, corresponding to the lower exchange
strength of 5.7~K reported for this compound\cite{manson2006}. The
data sets for both compounds have been analyzed by assuming the
specific heats have both lattice and magnetic contributions. The
lattice contribution in the low temperature region can be
approximated as $C_{\mathrm{lattice}}$=$\alpha T^3+\beta
T^{5}+\gamma T^7$, where $\alpha$, $\beta$, $\gamma$ are constants
to be determined. By modeling the Cu(Pz)$_2$(ClO$_4$)$_2$ data over
the entire range as a sum of the magnetic and lattice contributions,
we obtain the best fit for the parameters $J=17.7(3)$~K,
$\alpha=0.0039(2)$~J/K$^4\cdot$mol,
$\beta=-2.75\times$10$^{-6}$~J/K$^6\cdot$mol, and $\gamma=6.38
\times$10$^{-10}$~J/K$^{8}\cdot$mol, while the corresponding
parameters for [Cu(Pz)$_2$(HF$_2$)]BF$_4$ are $J=5.6(1)$~K,
$\alpha=0.0114(5)$~J/K$^4\cdot$mol,
$\beta=-6(1)\times$10$^{-5}$~J/K$^6\cdot$mol, and $\gamma=1.2(2)
\times$10$^{-7}$~J/K$^{8}\cdot$mol. The resultant fits to the two
compounds are shown as the solid lines while the estimated lattice
contributions for Cu(Pz)$_2$(ClO$_4$)$_2$ and
[Cu(Pz)$_2$(HF$_2$)]BF$_4$ appear as the lines of short dashes and
long dashes, respectively. The value of the exchange strength $J$ is
in excellent agreement with the value of 17.5~K obtained from
analysis of the magnetic susceptibility data \cite{turnbull,
albrecht, choi, woodward} and from the dispersion relation as
obtained by inelastic neutron scattering \cite{kenzelmann}. Likewise
the value of 5.6~K obtained for [Cu(Pz)$_2$(HF$_2$)]BF$_4$ agrees
well with the value of 5.7~K obtained from susceptiility
studies\cite{manson2006}.

\begin{figure}[h!]
\centering \subfigure[]{
\includegraphics[width=7.5cm]{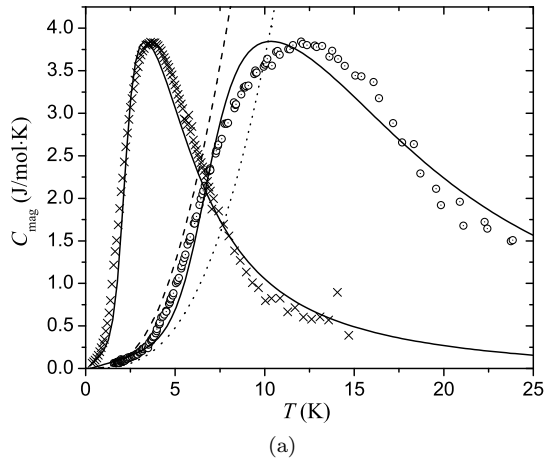}\label{Cmag_1}}
\subfigure[]{
\includegraphics[width=7.5cm]{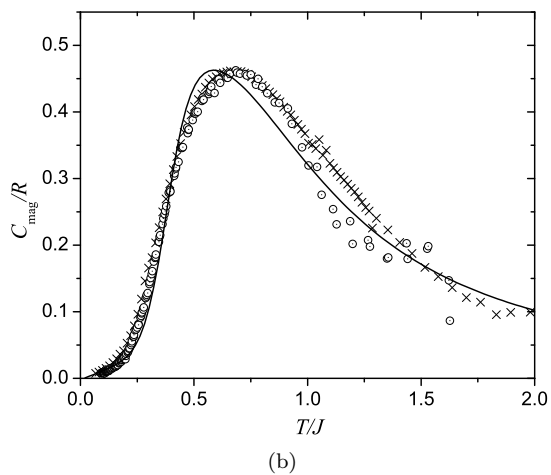}\label{Cmag_2}}
\caption{(a) Magnetic specific heats of Cu(Pz)$_2$(ClO$_4$)$_2$
($\odot$) and [Cu(Pz)$_2$(HF$_2$)]BF$_4$ ($\times$) as functions of
temperature with the solid lines corresponding to the best-fit
predictions for the magnetic specific heats based on the parameters
given in the text. The dashed lines (short dashes for
Cu(Pz)$_2$(ClO$_4$)$_2$ and long dashes for
[Cu(Pz)$_2$(HF$_2$)]BF$_4$) mark the estimated lattice contributions
that were subtracted from the data sets in 
Fig.~\ref{totalSpecificheat} to produce the magnetic specific heats. (b)
The reduced magnetic specific heats ($C_{\mathrm{mag}}/R$) of
Cu(Pz)$_2$(ClO$_4$)$_2$ ($\odot$) and [Cu(Pz)$_2$(HF$_2$)]BF$_4$
($\times$) as functions of the reduced temperature $T/J$ with the
solid line corresponding to the theoretical
prediction\cite{sengupta} for the magnetic specific heat of the
2DSLQHA.}
\end{figure}

The magnetic specific heats, obtained by subtracting estimated
lattice contributions from the total specific heats of 
Fig.~\ref{totalSpecificheat}, are shown in Fig.~\ref{Cmag_1}, along with
the estimated lattice contributions as the dashed lines, plus the
best fit to the theoretical prediction for the 2DSLQHAF. While the
overall agreement between data and theory is quite good, there are
systematic disagreements that are similar for each compound. The
theoretical prediction initially rises more slowly at low
temperature, surpasses the experimental data while reaching its
maximum value at a lower temperature, then decreases less rapidly
than the data at higher temperatures. Nevertheless, the entropy
changes for both data sets\cite{R_varied} and the theoretical curve
are all within several percent of $R\ln 2$. It is important to note
that neither data set shows an anomaly at the temperature of the 3D
ordering temperatures (4.21~K for Cu(Pz)$_2$(ClO$_4$)$_2$ [see below], 1.54~K
for [Cu(Pz)$_2$(HF$_2$)]BF$_4$ [see Ref.~\onlinecite{manson2006}]) 
as determined by $\mu^+$SR.

The comparison of experiment and theory is more clearly represented
in Fig.~\ref{Cmag_2}, in which the reduced specific heat
($C_{\mathrm{mag}}/R$) is plotted as a function of the reduced
temperature ($T/J$) for the theory and both experimental data sets.
Here it is noticed that the two experimental data sets are
essentially superimposible up to $T\approx J$. At higher
temperatures, the data for Cu(Pz)$_2$(ClO$_4$)$_2$ falls below that
of [Cu(Pz)$_2$(HF$_2$)]BF$_4$ and of the theory. However, by this
temperature, the lattice contribution of Cu(Pz)$_2$(ClO$_4$)$_2$ is
more than 90\% of the measured value, with commensurate
uncertainties in the magnitude of $C_{\mathrm{mag}}$.

\subsection{$\mu^{+}$SR}

\begin{figure}
\epsfig{file=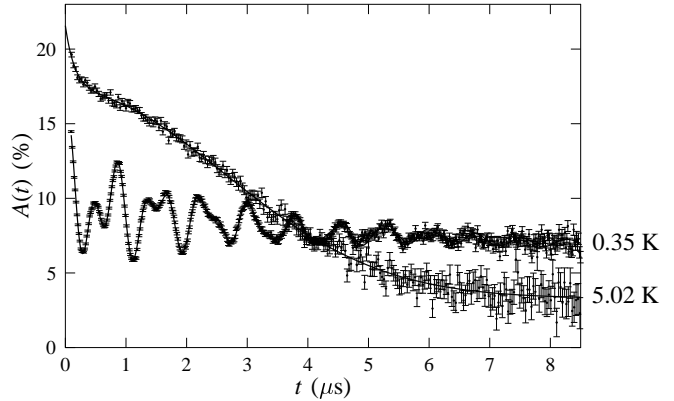,width=9cm} \caption{(a) ZF $\mu^{+}$SR spectra
in Cu(Pz)$_{2}$(ClO$_{4}$)$_{2}$
measured at $T=0.35$~K and 5.02~K. At temperatures $T \geq 4.27$~K,
Eq. (\ref{fitfunc2}) describes the data while the oscillations
measured below this temperature are fitted to Eq. (\ref{fitfunc}).
  \label{muonfig}}
\end{figure}

ZF $\mu^{+}$SR spectra measured on Cu(Pz)$_{2}$(ClO$_{4}$)$_{2}$ at
two temperatures are shown in Fig.~\ref{muonfig}.
For temperatures $T \geq 4.27$~K the measured spectra are found to contain two contributions.
The first is a fast relaxing
component $A_{1}$ which dominates the signal at early times and is
well described by an
exponential
function $\exp(-\lambda t)$.
The second is a larger, slowly relaxing component $A_{2}$, which
dominates at intermediate
times and fits
to the Kubo Toyabe (KT) function\cite{hayano} $f_{\mathrm{KT}}(\Delta, t)$
where $\Delta$ is the second moment of the static,
local magnetic field distribution
defined by $\Delta = \gamma_{\mu}
\sqrt{\langle (B - \langle B \rangle)^{2} \rangle}$,
where $B$ is the magnitude of the local magnetic field
and $\gamma_{\mu}$($=2 \pi \times 135.5$~MHz T$^{-1}$) is the muon gyromagnetic
ratio.
The KT function is characteristic of spin relaxation
due to a random, quasi-static distribution of local magnetic fields at
diamagnetic muon sites.
We do not observe the recovery in asymmetry at late times that is
expected for the static KT function.
The lack of this recovery is probably due to slow dynamics
in the random field distribution and is crudely modelled here with an exponential term\cite{francis}
$\exp(-\Lambda_{2} t)$.
The data were found to be best fitted with the resulting function
\begin{eqnarray}
A(t)&= &A_{1} \exp(-\Lambda_{1} t)
+A_{2}f_{\mathrm{KT}}(\Delta, t)\exp(-\Lambda_{2} t) \nonumber \\
& & +A_{\mathrm{bg}}\exp(-\lambda_{\mathrm{bg}}), \label{fitfunc2}
\end{eqnarray}
where $A_{\mathrm{bg}}\exp(-\lambda_{\mathrm{bg}} t)$ represents a constant background signal
from those muons that stop in the sample holder and cryostat tail.
The fitting parameters show very little variation in the range
$4.27 \leq T \leq 14$~K.
The component with amplitude $A_{1}$
(found to have relaxation rate $\Lambda_{1} \sim 10$~MHz)
very likely arises due the existence of a paramagnetic muon
state.
In the component with amplitude $A_{2}$, the magnitude of
$\Delta (\sim 0.2$~MHz) suggests that
the random magnetic field distribution giving rise to the KT function
is due to nuclear magnetic moments, implying that the field due to
electronic moments at these muon sites is motionally narrowed
out of the spectrum due to very rapid fluctuations.

In spectra measured at temperatures $T \leq 4.20$~K oscillations in the
asymmetry spectra are observed at several frequencies
(Fig.~\ref{muonfig}). These oscillations are characteristic of a
quasistatic local magnetic field at the muon stopping site, which causes a
coherent precession of the spins of those muons with a component of their
spin polarization perpendicular to this local field (expected to be 2/3 of
the total polarization). The frequency of the oscillations is given by
$\nu_{i} = \gamma_{\mu} B_{i}/2 \pi$, where $B_{i}$ is the magnitude of
the local magnetic field at the $i$th muon site. Any distribution in
magnitude of these local fields will result in a relaxation of the
oscillating signal, described by relaxation rates\cite{hayano}
$\lambda_{i}$. The presence of oscillations at low temperatures in
Cu(Pz)$_{2}$(ClO$_{4}$)$_{2}$ suggests very strongly that this material is
magnetically ordered below 4.20~K.

 Three separate
frequencies were identified in the low temperature spectra,
corresponding to three magnetically inequivalent muon sites in the
material.
The precession frequencies, which are proportional to the internal
magnetic field as experienced by the muon,
are proportional to the magnetic order parameter
for these system.
In order to extract the $T$-dependence of the frequencies,
the low temperature data were fitted to the functional form
\begin{eqnarray}
A(t) &=& A_{1} \exp (-\lambda_{1} t )\cos (2 \pi \nu_{1} t + \phi_{1})
\nonumber
\\
& &      +A_{2} \exp (-\lambda_{2} t )\cos (2 \pi \nu_{2} t + \phi_{2})  \nonumber \\
& &      +A_{3} \exp (-\lambda_{3} t )\cos (2 \pi \nu_{3} t +
\phi_{3}) \nonumber \\
& &      +A_{4} \exp(-\lambda_{4} t)+A_{\mathrm{bg}}\exp(-\lambda_{\mathrm{bg}} t)\label{fitfunc},
\end{eqnarray}
We note that phase offsets $\phi$ were required to fit the data (see
Table~\ref{fittable}) as observed in our previous $\mu^{+}$SR
studies of Cu$^{2+}$ systems \cite{tom1,tom2,manson}. The amplitudes
of the oscillating components were found to be constant in the
ordered state and were fixed at the values shown in Table
\ref{fittable}. These values suggest that the three muon sites have
occupation probabilities in the ratio $0.60:0.33:0.07$. The term
$A_{4} \exp(-\lambda_{4} t)$ accounts for the contribution from
those muons with a spin component parallel to the local magnetic
field expected to be half of the oscillating amplitude (see above).
From Table~\ref{fittable}, we see that $A_{4}/(A_{1}+A_{2}+A_{3})
\approx 0.45$ (i.e.\ the ratio of amplitudes resulting from local
magnetic field components parallel to the initial muon-spin direction 
to those perpendicular), is close to the expected value of 1/2, 
suggesting that the material is ordered throughout its bulk. Further 
evidence for the
presence of a magnetic phase transition is provided by the
observation that the relaxation rates $\lambda_{i}$ tend to increase
as $T_{\mathrm{N}}$ is approached from below, due to the onset of
critical fluctuations.

\begin{table}
   \begin{center}
\begin{ruledtabular}
   \begin{tabular}{llll}
$i$ &  $A_{i}$~(\%)  & $\phi_{i}$ \\
\hline
1   &  3.127      &   $-$39.5  \\
2   &  2.508     &    $-$21.88  \\
3   &  0.174     &   96.6   \\
4   &  2.633     &         \\
   \end{tabular}
\end{ruledtabular}
   \end{center}
   \caption{Fitting parameters for equation (\ref{fitfunc})
applied to data measured for $T \leq 4.20$~K.\label{fittable} }
\end{table}\

\begin{figure}
\epsfig{file=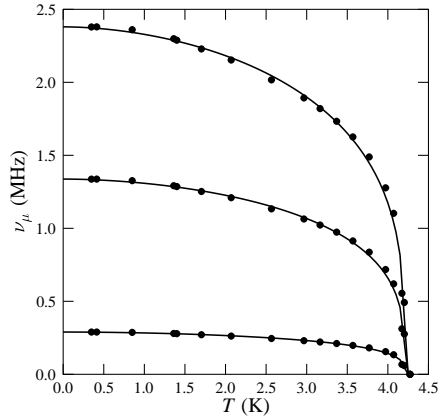,width=7cm} \caption{Temperature
evolution of the three magnetic oscillation frequencies
in Cu(Pz)$_{2}$(ClO$_{4}$)$_{2}$, extracted
from fits to Eq.(\ref{fitfunc}). The solid lines are fits to the
functional form $\nu_{i}(T)
=\nu_{i}(0)(1-(T/T_{\mathrm{N}})^{\alpha})^{\beta}$(see main text).
  \label{frequencies}}
\end{figure}

The three frequencies were found to be in the proportions
$\nu_{1}:\nu_{2}:\nu_{3} = 1:0.56:0.12$ and were fixed in this ratio
in the fitting procedure. The magnitude of these frequencies were
fitted as a function of temperature with parameters in
Eq.(\ref{fitfunc}) fixed at the values given in Table~\ref{fittable}
and relaxation rated allowed to vary. The resulting temperature
evolution of the precession frequencies is shown in
Fig.~\ref{frequencies}. From fits of the data to the form $\nu_{i}(T)
=\nu_{i}(0)(1-(T/T_{\mathrm{N}})^{\alpha})^{\beta}$, we estimate
$T_{\mathrm{N}}=4.21(1)$~K, $\alpha=1.8(3)$, $\beta = 0.29(2)$,
$\nu_{1}(0)=2.38(3)$~MHz, $\nu_{2}(0)=1.33(2)$~MHz and
 $\nu_{3}(0)=0.29(2)$~MHz.

\section{Discussion}

A method for estimating the interlayer coupling constant $J'$ in 3D
arrays of 2DSLQHA has recently been developed, based on a modified
random phase approximation, modelled with classical and quantum
Monte Carlo simulations \cite{yasuda}. This approach leads to an
empirical formula relating $J'$ and $T_{\mathrm{N}}$,
\begin{equation}
|J'/J|= \exp \left( b-\frac{4 \pi \rho_{\mathrm{s}}}{T_{\mathrm{N}}}
\right),
\label{jprime}
\end{equation}
where $\rho$ is the spin stiffness, given \cite{beard} by
$\rho_{\mathrm{s}}=0.183 J$ and $b=2.43$ for $S=1/2$. Using
$J=17.7$~K, we obtain $\rho_{\mathrm{s}}=3.26$~K and $|J'/J|=6.8
\times 10^{-4}$. Table~\ref{comptable} compares these parameters to
those for other 2D layered compounds. We see that
Cu(Pz)$_{2}$(ClO$_{4}$)$_{2}$ compares very favourably as a highly
anisotropic system, with the layers around an order of magnitude
better isolated than those in Ca$_{0.85}$Sr$_{0.15}$CuO$_{2}$,
although not as successful a realisation of a 2DSLQHA as
Sr$_{2}$CuO$_{2}$Cl$_{2}$, where Eq.(\ref{jprime}) yields $|J'/J|
\sim 10^{-5}$.The $J'/J$ ratios were obtained assuming perfectly
Heisenberg exchange interactions. The presence of any exchange
anisotropy will also raise the ratio $T_N/J$ so by ignoring the
ansiotropy fields while using Eq.(\ref{jprime}), we are setting an
upper bound for the $J'/J$ ratios.

Cu(Pz)$_2$(ClO$_4$)$_2$ and [Cu(Pz)$_2$(HF$_2$)]BF$_4$ contain
structurally similar copper pyrazine layers separated either by the bulky
perchlorate groups or by the HF$_2$ anions. Their intralayer exchange
strengths differ by a factor of three (17.7~K, 5.6~K) but their low ratios
of $T_N/J$ show them to be well isolated ($J'/J\sim$ 10$^{-3}$, see
Table~\ref{comptable}). As seen in Fig.~\ref{Cmag_2}, their magnetic
specific heats are essentially identical up to relative temperatures of
$T\approx J$. Nevertheless, there is noticable difference between their
behavior and that predicted for isolated 2DSLQHA model
systems\cite{sengupta}. This difference may arise from the small terms in
the Hamiltonian relevant for the compounds but ignored in the Monte Carlo
simulation. Studies of Cu(Pz)$_2$(ClO$_4$)$_2$ in the magnetically ordered
state show the presence of about a 0.3~T axial field perpendicular to the
magnetic layers \cite{xiao}. While small compared to the 60~T exchange field,
this axial field will break the rotational symmetry of the Heisenberg
Hamiltonian and will modify the low-temperature behavior. It will be
necessary to examine the specific heat of the quasi-2DSLQHA in the
presence of both weak interlayer interactions as well as weak anisotropies
to understand these effects qualitatively.

\begin{table}
   \caption{Parameters for
the  layered compounds 
Sr$_{2}$CuO$_{2}$Cl$_{2}$ \cite{greven},
Cu(Pz)$_{2}$(ClO$_{4}$)$_{2}$ \cite{choi},
[Cu(Pz)$_{2}$(HF$_2$)]BF$_4$\cite{manson2006},
Ca$_{0.85}$Sr$_{0.15}$CuO$_{2}$ \cite{siegrist},
(5MAP)$_{2}$CuBr$_4$ \cite{woodward2} with values of $|J'/J|$
estimated from Eq.(\ref{jprime}).\label{comptable} }
     \begin{ruledtabular}
   \begin{tabular}{lllll}
&  $|J|/k_{\mathrm{B}}$~(K)& $T_{\mathrm{N}}$~(K) &
$|k_{\mathrm{B}}T_{\mathrm{N}}/J|$ & $|J'/J|$ \\\hline
Sr$_{2}$CuO$_{2}$Cl$_{2}$ & 1451 & 256.5 & 0.18 & $2.5 \times 10^{-5}$ \\
Cu(Pz)$_{2}$(ClO$_{4}$)$_{2}$& 17.7 & 4.21 & 0.24 & $6.8 \times 10^{-4}$\\
$[$Cu(Pz)$_{2}$(HF$_2$)]BF$_4$ & 5.7 &1.54 &0.27 &$ 2.3\times 10^{-3}$\\
Ca$_{0.85}$Sr$_{0.15}$CuO$_{2}$ & 1535 & 537 &  0.35  & $1.6 \times 10^{-2}$\\
(5MAP)$_{2}$CuBr$_{4}$&  6.5 & 3.8 & 0.58 & $2.1 \times 10^{-1}$\\
   \end{tabular}
     \end{ruledtabular}
\end{table}

\section{Conclusion}

We have examined the 2D Heisenberg antiferromagnet
Cu(Pz)$_{2}$(ClO$_{4}$)$_{2}$ using specific heat and  $\mu^+$SR
techniques. The value of the exchange strength obtained ($J$ =
17.7(3)~K) from the magnetic specific heat is in excellent agreement
with values obtained from susceptibility \cite{turnbull, albrecht,
choi, woodward} and neutron scattering \cite{kenzelmann}
experiments. The muon spin relaxation studies demonstrated the
existence of 3D long range magnetic order throughout the bulk
of the sample in
Cu(Pz)$_{2}$(ClO$_{4}$)$_{2}$ at temperatures below 
$T_{\mathrm{N}}$ = 4.21(1)~K. The
value of $T_{\mathrm{N}}/J$ (0.24) corresponds to a very low ratio
of interlayer to intralayer exchange strengths
($|J'/J|$=6.8$\times$10$^{-4}$). The combination of good isolation
of the magnetic layers plus relatively small exchange strength makes
Cu(Pz)$_{2}$(ClO$_{4}$)$_{2}$, as well as
$[$Cu(Pz)$_{2}$(HF$_2$)]BF$_4$, good candidates for studies of the
field dependence of the energy spectrum of the 2DSLQHA
\cite{syljuasen, zhitomirsky}.

The magnetic ordering transition is readily apparent in the $\mu^+$SR data
(Fig.~\ref{muonfig}) even though no anomaly is apparent in the specific
heat results at that temperature or at $T_N$ of
[Cu(Pz)$_{2}$(HF$_2$)]BF$_4$ (Fig.~\ref{Cmag_1}). The absence of an
ordering peak in the specific heat for well isolated low-dimensional
magnets was previously anticipated \cite{sengupta} in Monte Carlo studies.
These demonstrate that most of the available spin entropy ($R\ln 2$ per
mole) is associated with the development of in-plane correlations which
give rise to the broad peak in $C_{\rm mag}$ centred around $T=0.65 J$.  This
leaves little remaining entropy to be associated the 3D ordering at
$T=T_{\rm N}$. The critical temperature of Cu(Pz)$_2$(ClO$_4$)$_2$ was
also not detected by previous inelastic neutron scattering experiments on
polycrystalline samples; only after adequate single crystals became
available could the transition be observed with this
technique\cite{kenzelmann}. These results emphasize the value of muons in
detecting magnetic order in polycrystalline materials.

\begin{acknowledgments}
  We are grateful to Philip King at ISIS for technical assistance.  This
  work is supported by the EPSRC (UK) and by an award from
  Research Corporation (USA). T.L.\ acknowledges support from the
  1851 Commission (UK). F.A.C., S.S. and M.A.N.  are grateful for the
  financial support from CNPq, CAPES and FAPERJ (Brazil). 
 Work at Argonne National Laboratory was supported by the Office of Basic
 Energy Sciences, Division of Materials Science, U.S. Department of
 Energy under contract no.\ DE-AC02-06CH11357.
 We thank Pinaki
  Sengupta for the use of his QMC simulation results.
\end{acknowledgments}


\begin{thebibliography}{xx}
\bibitem{R_varied}
It was necessary to increase all the specific heat data of
[Cu(Pz)$_2$(HF$_2$)](BF$_4$) by 15\% to generate an entropy change
of $R\ln 2$ and to match the amplitude of the theoretical specific
heat.
\end{thebibliography}
\end{document}